\documentclass[%
 aip,
 jmp,%
 amsmath,amssymb,
 reprint,%
]{revtex4-2}

\usepackage{graphicx}
\usepackage{dcolumn}
\usepackage{bm}

\preprint{ }
 \usepackage[version=4]{mhchem}

\usepackage[normalem]{ulem}
\usepackage{graphicx}
\usepackage{dcolumn,xcolor}
\usepackage{bm}
\begin{document}
\preprint{}

\title[Surface Analysis of HEA]{Nanoscale Surface Analysis of High Entropy Alloy}

\author{
Hsin-Hui Huang$^{1,2,3*}$, Meguya Ryu$^{4*}$, Yoshiaki Nishijima$^{5,6,7}$, Haoran Mu$^{1,3}$, Mohit Kumar$^8$, Nguyen
Hoai An Le$^1$, Adrian Cernescu$^9$, 
 Jitraporn Vongsvivut$^{10}$, Andrew Siao Ming Ang$^2$, Lina Grineviciute$^{11}$, Challapalli Subrahmanyam$^8$, Junko Morikawa$^{4,12,13}$, Saulius Juodkazis$^{1,12,14}$
}%

\affiliation{Optical Sciences Centre, 
Swinburne University of Technology, Hawthorn, Victoria 3122, Australia}
\affiliation{Australian Research Council (ARC) Industrial Transformation Training Centre in Surface Engineering for Advanced Materials (SEAM), Swinburne University of Technology, Hawthorn, VIC, 3122, Australia}
\affiliation{Melbourne Centre for Nanofabrication (MCN), 151 Wellington Road, Clayton, Vic 3168, Australia}
\affiliation{School of Materials and Chemical Technology, Institute of Science Tokyo, 2-12-1, Ookayama, Meguro-ku, Tokyo 152-8550, Japan}
\affiliation{Institute of Advanced Sciences, Yokohama National University, 79-5 Tokiwadai, Hodogaya-ku, Yokohama, 240-8501, Kanagawa, Japan}
\affiliation{Institute for Multidisciplinary Sciences, Yokohama National University, 79-5 Tokiwadai, Hodogaya-ku, Yokohama, 240-8501, Kanagawa, Japan}
\affiliation{PRESTO, JST, Japan}
\affiliation{Department of Chemistry, Indian Institute of Technology Hyderabad, Sangareddy, Telangana, 502285, Kandi, India}
\affiliation{Attocube Systems AG, Haar-Munich, Germany}
\affiliation{Infrared Microspectroscopy (IRM) Beamline, ANSTO‒Australian Synchrotron, 800 Blackburn Road, Clayton, Victoria 3168, Australia}
\affiliation{Center for Physical Sciences and Technology, Savanoriu ave. 231, LT-02300,Vilnius, Lithuania}
\affiliation{World Research Hub (WRH), School of Materials and Chemical Technology, Institute of Science Tokyo, 2-12-1, Ookayama, Meguro-ku, Tokyo 152-8550, Japan}
\affiliation{Research Center for Autonomous Systems Materialogy (ASMat), Institute of Innovative Research, Institute of Science Tokyo,Yokohama 226-8501, Japan}
\affiliation{Laser Research Center, Physics Faculty, Vilnius University, Saul\.{e}tekio Ave. 10, 10223 Vilnius, Lithuania}


\thanks{*Correspondence: H-H. H. hsinhuihuang@swin.edu.au; M. R. ryu.meguya@mct.isct.ac.jp  }

\date{\today}

\begin{abstract}
Nanoscale surface analysis of $\sim 1~\mu$m thick high entropy alloys (HEAs) was carried out using nano-IR for hyperspectral imaging and single point spectroscopy in the 700-1700~cm$^{-1}$ spectral range. Nano-IR is based on the detection of scattered light from an oscillating metal coated nano-tip in one of the arms of the Fourier transform infrared spectrometer and has a resolution defined by the tip radius of the probe, $\sim 20$~nm, regardless of the excitation wavelength. HEA CuPdAgPtAu showed an absorption and reflection increase at 900-1100~cm$^{-1}$ band, which is consistent with Drude-Lorenz modeling of permittivity, however, could also signify oxide formation as tested by X-ray photoelectron spectroscopy of CuPdAgPtAu and CrFeCoNiCuMo. 
Realization of polarization analysis for nano-IR nano-spectroscopy in the plane perpendicular to the sample's surface is discussed and modeled. The currently available modality of surface analysis with the excitation-detection mode of the p-pol. antenna can be extended to full 3D analysis of the orientational dependencies of local absorbance and refractive index.     
\end{abstract}

\maketitle


\tableofcontents

\section{Introduction}

The nanoscale Fourier transform infrared (nano-FTIR) provides IR spectroscopy at a spatial resolution of atomic force microscopy (AFM), delivering nanoscale chemical identification at the spectral region of the materials' fingerprint, 
hyperspectral imaging, and characterization of the structure of catalytic materials~\cite{10,YANG,Matthieu,Law,Gupta,2dsilk}. 
Nanoscale characterization of materials, for example, metal-dielectric phase Mott transitions~\cite{McL,Mott}, observation of surface plasmons on graphene~\cite{Frank,Fei}. FTIR based on synchrotron radiation provides high brightness IR beams which can be coupled with nano-tip scanning spectroscopy~\cite{Freitas,xe,Herr,omar,Santos,sins}. 

High entropy alloys (HEAs) are currently an active research topic due to their superior and tunable mechanical properties for applications in high-pressure / temperature environments encountered in specialty metal production, nuclear power stations, defense, and space applications~\cite{Hui25,birb,Ali,Ali1}. It is well known from the material science of complex composites and alloys that the exact nanoscale properties have to be directly measured and defined as input to computational finite element methods to \emph{quantitatively} predict mechanical, thermal, and optical properties~\cite{ZAMENGO,cal} rather than relying on \emph{a qualitative} description of the underlying physical mechanisms. In this regard, the characterization of HEAs on nano- and micro-scales is expected to bring about the required detailed understanding of properties, their local distributions, and compositional changes, which are critically important for the aforementioned applications. Broadband stealth of HEA surfaces via the IR-to-THz wavelengths is expected due to low reflectivity~\cite{steal}. The CrFeCoNiCuMo 
HEA nano-films were used to determine the complex optical refractive index $(n+i\kappa)$ and to form micro-droplets by laser-induced forward transfer (LIFT) printing~\cite{22m8063}. HEA of noble metals CuPdAgPtAu can be used to engineer meta-surfaces for perfect absorbers, hence perfect emitters, at the designed IR wavelengths at which the scattering cross sections of absorbance and scattering become equal~\cite{25e81}.        

For nanoscale analysis, the sub-diffraction scattering is based on a metalized AFM tip~\cite{12} and was adopted in this study. The tip 
acts as a light-concentrating antenna probing the sample surface topography with a nano-focused light field under external IR illumination provided by a broadband difference-frequency generation laser (DFG; Toptica). The tapping mode operation with $\sim$60~nm amplitude modulates the near-field interaction between the tip and the sample. The measured nano-IR absorbance is proportional to the imaginary part of the scattering coefficient $\sigma_n(\omega) = s(\omega)e^{-i\phi(\omega)}$, where $s(\omega)$ and $\phi(\omega)$ are the amplitude and phase of the back-scattered spectra. The light scattered field $E_s(\omega)$ is related to the incident field $E_i(\omega)$ through the equation $E_s=\sigma_sE_i$~\cite{10}. With the asymmetric Michelson interferometer, where AFM tip/antenna is in one of its arms, the full complex function of the scattered optical signal can be recorded, therefore enabling the simultaneous measurement of both nano-IR absorbance and reflectivity spectra~\cite{10}. The interferometer with lock-in detection of the signal at the higher harmonic of the tapping frequency $\sim$250~kHz provides background-free nano-IR spectra and images with maximum resolution defined by the AFM tip size independent of the laser wavelength~\cite{10}. Different harmonics also add depth probing capability. This technique previously tested on microtome sliced silk~\cite{19as3991,19bjn922} to reveal optical anisotropy for two perpendicular slices was applied to HEAs in this study. Anisotropy in alignment of polymer brushes can be resolved via nano-IR mapping~\cite{orien}. 

Here, top- and side-view imaging of HEA micro-films is presented at the chemical finger printing spectral range using nano-IR. Such characterization can help to explain the performance of HEA micro-films in high pressure experiments using optical Joule-energy level pump pulses and a sub-10~fs X-ray free-electron laser (XFEL) probe for dynamic structural analysis~\cite{arX} 
The extension of nano-IR into the characterization of orientation and anisotropy at nanoscale is discussed using a toy-model numerical modeling with finite different time domain (FDTD) method. It proves that current 4-pol. anisotropy characterization at Australian Synchrotron IR microscopy beamline could be extended into nanoscale anisotropy measurements at the sub-surface regions of the sample. 

\section{Experimental: samples and methods}

\subsection{Characterisation techniques}

IR-NeaSCOPE$^{\mathrm{+s}}$ was used for nano-FTIR measurements. The principle of operation is based in recording the interferogram on the sample and reference (Au, Si, etc.) using a difference frequency generation (DFG) laser source. Applying a Fourier transform (FT) yields the local nano-IR spectrum. By normalizing the sample spectrum to that of a reference, the instrumental spectral response is eliminated, providing the optical Amplitude and Phase related to the Real part of FT (Reflectance) and Imaginary (Absorbance). Nano-IR absorption and reflectivity properties are simultaneously recorded with AFM topography and nano-local mechanical property - the stiffness.

The soft tapping operation mode of the IR-NeaSCOPE$^{\mathrm{+s}}$ microscope was used for high sensitivity 
and high resolution $\sim 10$~nm optical measurements at 
low laser power and short measurement times without alteration of the investigated 
materials (polymer matrix was used for embedding samples before microtome slicing and black Kapton was used as substrate for the deposition of the HEA coating), samples of the same thickness in black Kapton were used in nano-IR experiments as in XFEL~\cite{arX}.


X-ray photoelectron spectroscopy (XPS) was used for surface chemical characterization of the HEA samples using an AXIS Supra (Kratos Analytical) spectrometer equipped with a monochromatic Al K$_\alpha$ X-ray source (1486.6~eV) and a monoatomic Ar$^+$ ion gun. XPS spectra were acquired both before and after Ar$^+$ sputter etching for 2 and~4 min in slot mode to investigate surface and near-surface compositional variations with a spot size of 20~$\mu$m for analysis. XPS provides information on the elemental composition, chemical states, and environment of the material within an analysis depth of approximately $\sim$10~nm.

\subsection{Sample preparation}

Au-HEA samples were prepared by magnetron co-sputtering from CuAgAu and PdPt targets on cover glass and black Kapton substrates (MPS-6000, ULVAC). The sputtering target was centered over the direction to the substrate holder, and to maintain homogeneous deposition, substrates were rotated during sputtering. In addition, the substrates were annealed up to 150$^{\circ}$C to mix the alloy as established in previous studies~\cite{25e81}. 

Magnetic Fe-HEA was prepared by, first, thermal spraying of micro-powders of constituent materials and, then, coated on black Kapton by magnetron sputtering (AXXIS, JK Lesker). The final composition was close to the equal portions of six metals CrFeCoNiCuMo, where Cu came from the backing substrate during sputtering. Fe-HEA deposition onto a 3-mm thick copper plasma-sputtering target backing plate was described elsewhere~\cite{22m8063}. In short, commercially available gas-atomized (GA) CrFeCoNiMo 
HEA powder (Jiangsu Vilory, Xuzhou, China) with a particle size between 15 and 53~$\mu$m was used as feedstock to fabricate the HEA coating. The CrFeCoNiMo 
powder was sprayed using a high velocity oxygen fuel (HVOF) thermal spray system (GTV HVOF K2, GTV Verschleißschutz GmbH, Luckenbach, Germany). The spray parameters were: kerosene fuel had a flow rate of 28 liters/min, the \ce{O2} flow rate was 950~liters/min, the stand-off distance was 380~mm, and the powder feed rate was 50~g/min. The carrier gas was Ar at a flow rate of 7~liters/min. The surface of the HEA sputtering targets had a roughness of up to 10~$\mu$m.

\begin{figure*}[tb]
\centering\includegraphics[width=1\textwidth]{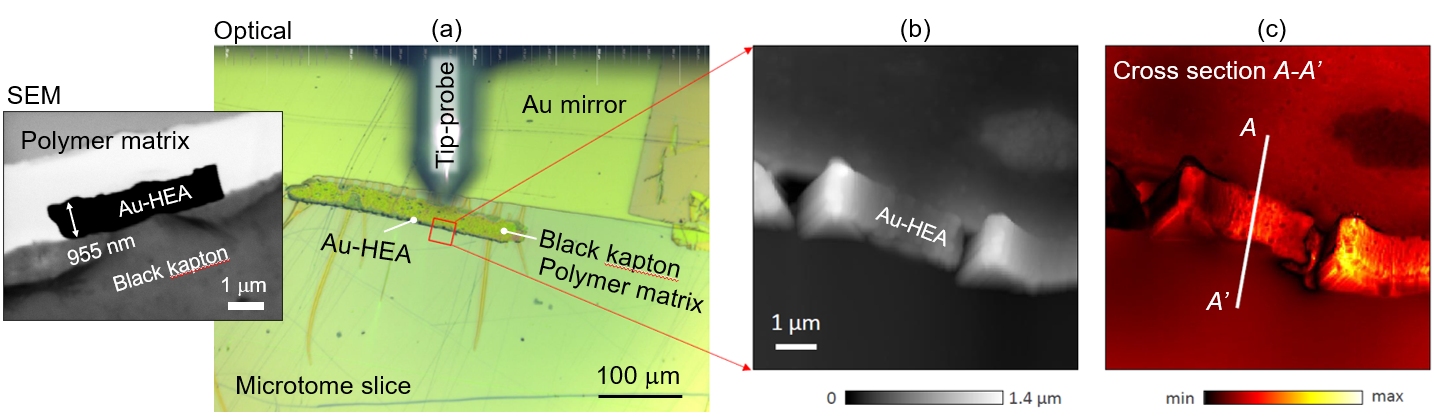}
\caption{\label{f-slic} (a) Optical image of a side-view microtome cut Au-HEA sample on black Kapton. Thickness of the microtome slice $960\pm 40$~nm (see SEM inset). Resolution of optical micro-image 0.7~$\mu$m. (b) Topography (height) from $8\times 8~\mu$m$^2$ in $200\times 100$~pixels, 9~ms/pixel. (c) IR optical signal (spectrally integrated); see Fig.~\ref{f-AR} for the reflectance and absorbance maps and averaged spectra along \emph{A-A'} image cross section.
}
\end{figure*}

The 90-nm thick microtome slices were made using Leica EM UC6. HEA samples on 25-$\mu$m of black Kapton (DuPont\texttrademark Kapton B) were embedded into polymer matrix (acrylate type light curable resin, Aronix LCR D-800, Toagosei co. Ltd., Tokyo, Japan), which has main spectral features at 1750~cm$^{-1}$ (carbonyl group), 1159~cm$^{-1}$ (ester C-O-C), 995~cm$^{-1}$ (C-O stretching), and around 900~cm$^{-1}$ due to formation of vinyl group~\cite{ALLEN19979}. The same procedure was applied for a hard and brittle tooth's dental tissue with enamel and was proven not to introduce IR spectral artifacts when polarisation-resolved synchrotron-IR micro-spectroscopy was performed~\cite{25nse70099}. 

Even thinner samples are required for high-resolution transmission electron microscope (TEM) to image the crystalline structure of HEA; however, this was beyond technical capability of the used microtome. Ion milling (typically Ga) based method is used for such lamella preparation but was not selected due to inevitable Ga contamination of metallic HEA samples.

\subsection{Numerical modeling by Finite Difference Time Domain (FDTD)}


The computational domain near the tip-to-surface plane (approximately $2.5 \times 3\times 2~\mu$m$^3$) was terminated by perfectly matched layers (PML) in all directions. A non-uniform mesh was employed, with a local mesh override of 2~nm applied to the nano-cylinder, the 20~nm air gap, and the upper portion of the substrate to accurately resolve strong near-field gradients. Outside of this region, the mesh was automatically adapted to reduce computational cost. Conformal mesh refinement was enabled to accurately represent the curved metallic boundaries of the 20~nm diameter Au nano-cylinder and to minimise stair casing errors. Numerical convergence was verified by reducing the mesh size until variations in peak near-field enhancement were below $3~\%$. Focused illumination ($NA = 0.95$) was implemented using a built-in Gaussian beam source with thin-lens focusing, with the focal plane positioned at the apex of the Au nano-cylinder. The beam was incident along the y-direction. Field monitors were positioned in the XZ-plane, perpendicular to the propagation direction (y-axis), to extract cross-sectional intensity distributions of the tangential (t) $E_x$ and normal (n) $E_z$ electric field components.

The wavelength-dependent optical constants of Au were taken from Johnson and Christy as implemented in the Lumerical material database. The refractive index of Si was obtained from the Palik database. The refractive index of \ce{Al2O3} was imported from the refractive index database of Lumerical, Ansys. 

\begin{figure*}[b!]
\centering\includegraphics[width=1\textwidth]{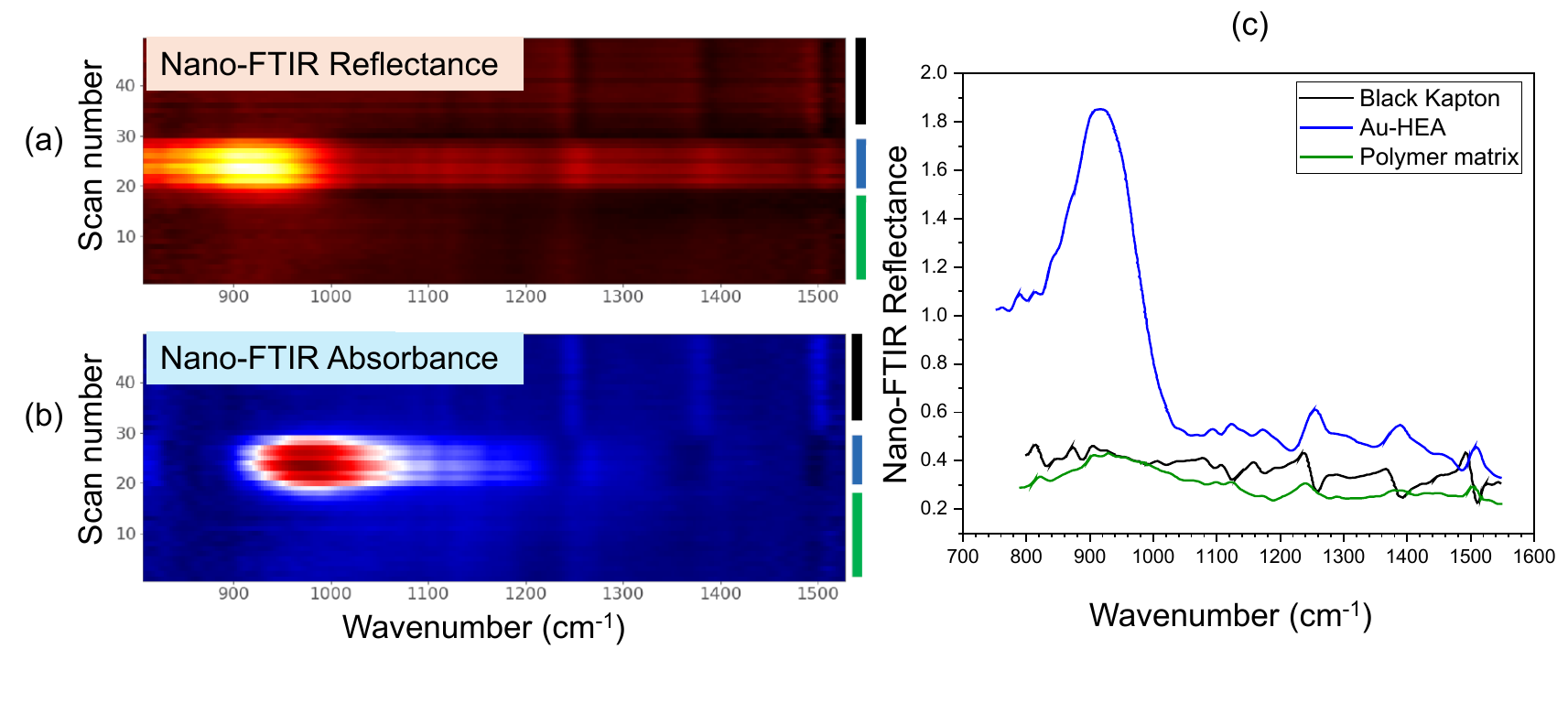}
\caption{\label{f-AR} Spectral mapping: 50 spectra recorded across 5 $\mu$m distance; spectral resolution: 12.5 cm$^-1$, measurement time 42 minutes. (a) Nano-FTIR reflectivity. (b) Nano-FTIR absorption. (c) Reflectivity averaged over three materials: black Kapton (top in (a,b)), Au-HEA (middle), microtome polymer matrix (bottom).    }
\end{figure*}

\section{Results}

Figure~\ref{f-slic} shows geometry of 90-nm-thick microtome slice of the Au-HEA sample on black Kapton (25~$\mu$m) for side view imaging by nano-IR. The slice of Au-HEA had sub-$1~\mu$m surface roughness. A flatter region was selected for IR spectral characterisation along the \emph{A-A'} line (Fig.~\ref{f-slic}(c)), which crossed the sample from the region of black Kapton substrate (polyimide), then throughout the Au-HEA, and into the matrix polymer, which was used to embed the sample for microtome slicing.    

\begin{figure*}[b]
\centering\includegraphics[width=1\textwidth]{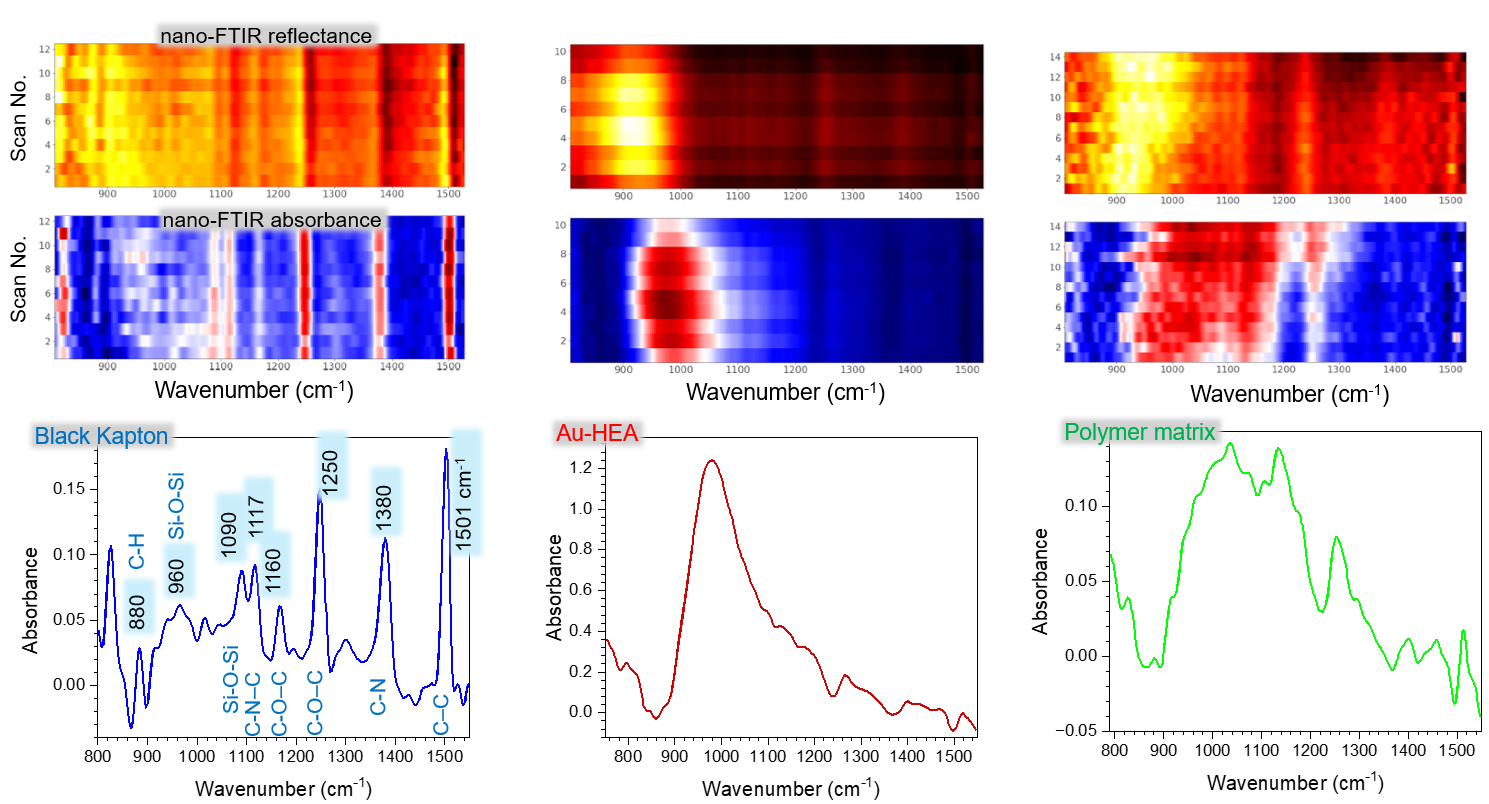}
\caption{\label{f-spec} Maps of nano-FTIR reflectance, absorbance, and averaged absorbance for: black Kapton (left), Au-HEA (middle), microtome polymer matrix (right). The black-Kapton bands are referenced well to polyimide with recognisable Si-O-Si (or Si-OH) bands related to silica filler.   }
\end{figure*}

Figure~\ref{f-AR} shows spectral maps of nano-FTIR reflectance $R$ and absorbance $A$ across 5~$\mu$m width scanned in 50 lines (step between scans is 100~nm). Clear spectral distinctions appear in the $R,A$ maps, which are presented as average along the selected three materials in Fig.~\ref{f-AR}(c). Spectra of $A,R$ and averaged absorbance for three separate materials is shown in Fig.~\ref{f-spec}. A clear polyimide (basis for black Kapton) is recognisable with matching characteristic peaks observed in polyimide sandwiched between Au-mirror and Au-nanodisks metamaterial~\cite{22jmcc451}. Polyimide has several bands with an orientation in the flat monomer. Those bands are used for determination of molecular alignment: 1780~cm$^{-1}$ the C=O symmetric (chain axis), 1745~cm$^{-1}$ C=O asymmetric (ring tilt), 1520~cm$^{-1}$ \ce{C6H5} aromatic rings (phenyl) tangential vibrations, and 1400 cm$^{-1}$ C-N-C axial stretch~\cite{21as1544}. 
The parallel bands to the polyimide film are 1491~cm$^{-1}$ for the skeletal vibration of the C-C of the 1,2,4-trisubstituted benzene ring in the diamine moieties and 1501~cm$^{-1}$ C-C. This is skeletal vibration of 1,4-di-substituted benzene in diamine moieties, while the perpendicular band is at 1724~cm$^{-1}$ for the C=O out-of phase asymmetric stretching of carbonyl bonds in imide rings. 

The polymer matrix used for the embedding of HEA is a water-borne acrylic polymer whose characteristic IR absorption band can be recognized. The absorption band from 900~cm$^{-1}$ to 1200~cm$^{-1}$ can be assigned to known absorption bands at 1159~cm$^{-1}$ (ester C-O-C), 995~cm$^{-1}$ (C-O stretching), and around 900~cm$^{-1}$ due to formation of vinyl group during photo-polymerization~\cite{ALLEN19979}. A prominent absorbance region with a peak at 920~cm$^{-1}$ and FWHM $\sim 200$~cm$^{-1}$ is typical for oxides. The Au-HEA is composed of noble metals, however, some of them can be more prone to oxidise into their native oxide forms: Ag, Cu. For example it is present in metal rich minerals such olivine \ce{(MgFe)2SiO4} and spinel group minerals of the structure \ce{XY2O4}, e.g., the spinel \ce{MgAl2O4}~\cite{Harr}. 

\begin{figure*}[tb]
\centering\includegraphics[width=1\textwidth]{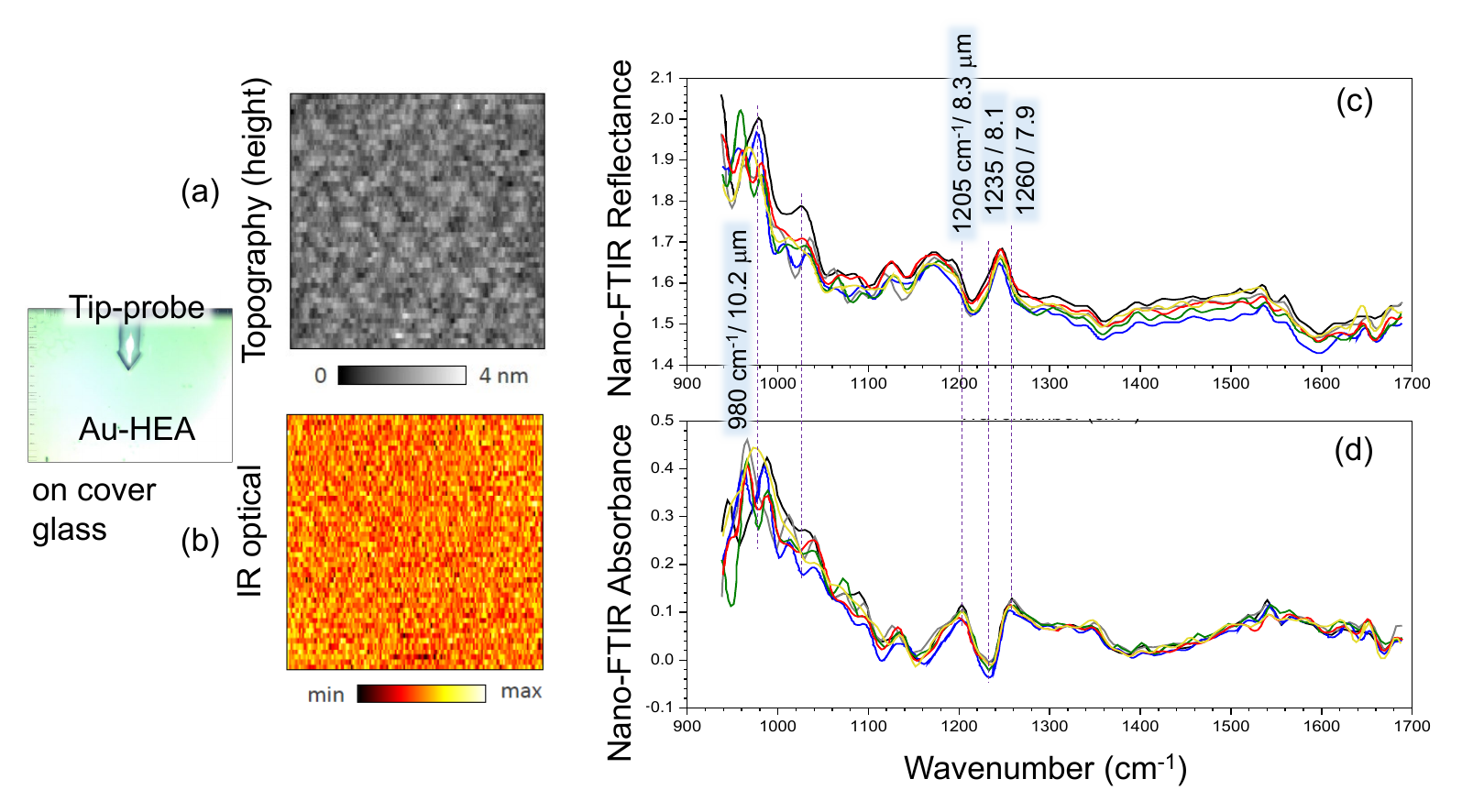}
\caption{\label{f-mirr} Topography (height) and IR optical maps: $5\time 5~\mu$m$^2$, $150\times 150$~pixels, 9~ms/pixel. Nano-FTIR reflectance and absorbance at six random locations (normalised to Si). Sample: cover glass coated with Au-HEA. Experiment: 10 averaged interferograms per measurement point; 1024~pixels per interferogram, 10~ms per pixel, spectral resolution 12.5~cm$^{-1}$.
  }
\end{figure*}

For further down-sizing the measurement region, single point $R,A$ spectra were measured from Au-HEA coated on cover glass; thickness of mirror was $\sim 1~\mu$m. The smooth nanoscale topography was recorded and the measurements of a single point were taken at six locations with averaging 10 times (Fig.~\ref{f-mirr}). Characteristic of Au-HEA $R$ and even more uniform $A$ spectra were obtained at random points in the area with a few micrometers in cross section. The separation between single points was greater than that between 100~nm line scans. The main spectral feature is the increase in $R$ and $A$ (less) in the 970-1100~cm$^{-1}$ range, which is consistent with the Drude-Lorenz model of permittivity for this composition of Au-HEA~\cite{25e81}.

\begin{figure*}[tb]
\centering\includegraphics[width=.9\textwidth]{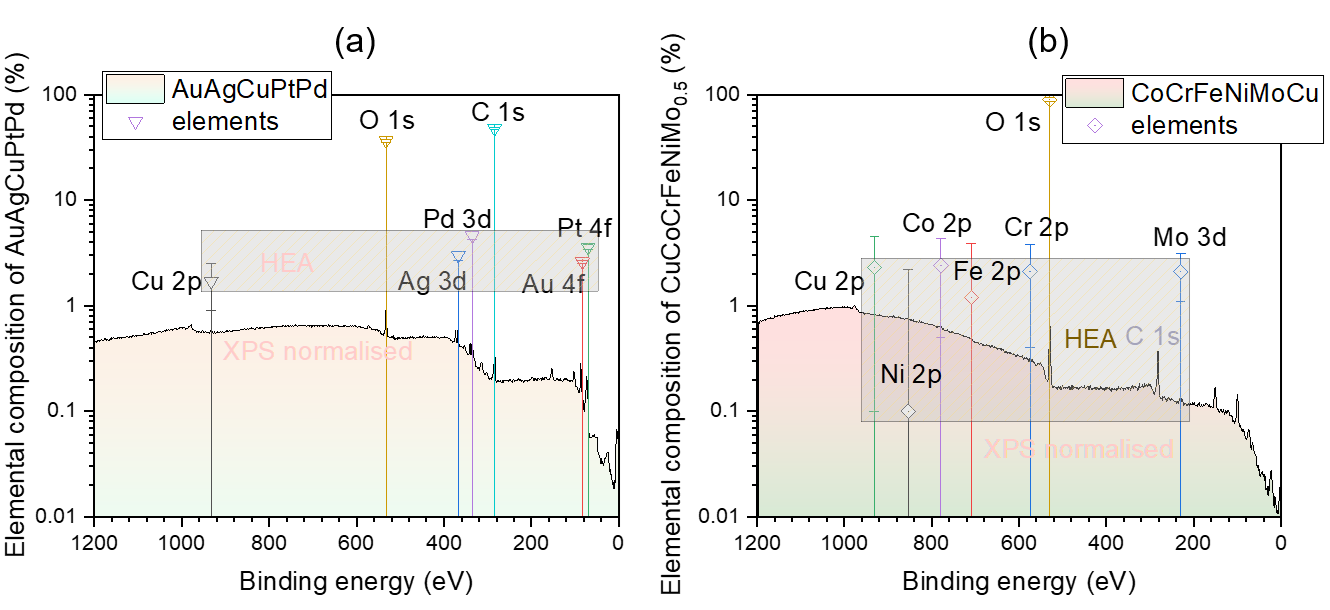}
\caption{\label{f-xps} 
X-ray photo-electron spectrum (XPS) analysis of two HEA samples on black Kapton: (a) Au-HEA \ce{CuPdAgPtAu} and (b) Fe-HEA CrFeCoNiCuMo. 
Carbon was not taken into account for elemental atomic concentrations in (a); it is not shown in (b). Au-HEA and Fe-HEA were Ar plasma cleaned for 2~min and 4~min, respectively before quantification measurement was taken.    }
\end{figure*}

XPS 
was performed at a 20~$\mu$m diameter area on black Kapton coated Au-HEA and Fe-HEA (Fig.~\ref{f-xps}). Oxygen was a prominent feature in XPS spectra. The spread of elemental percentages for Au-HEA was narrower compared to Fe-HEA. This is consistent with the method of preparation of HEA by magnetron sputtering. Fe-HEA was made by thermal spray and oxidation is more expected as compared with sputtering targets of pure metallic content CuAgAu and PdPt, which were used for Au-HEA sputtering from targets with fixed stoichiometric mixture. 

\section{Discussion}

Next, a numerical study is presented to show the possibility of polarisation analysis using nano-probe. 

\subsection{Nano-rough HEA for low reflectance and absorbance} 

Near-IR absorbance and reflectance $A,R$ can be determined from the normal $E_z$-field (non-propagating near-field) component to the sample's surface. The dipole interaction induced by antenna depends on the intensity along the z-direction as $|E_z|^2\propto 1/z^6$. The $A$ and $R$ components in this direction are probing corresponding surface and subsurface optical properties defined by nano-local refractive index $(n+i\kappa)$ (Fig.~\ref{f-toy}(a)). The scattered light is radially-perpendicular, i.e. in directions $x,y$ to that of the dipole (z-axis). The scattered E-field has the same orientation of polarisation $E_z$ as observed from the dipole site (Fig.~\ref{f-toy}). Hence, when surface is nano-rough and the roughness is on the scale of the dipole interaction (a tip size), the surface is subjected to the tip-scattered field. This scattered field has a range of incidence angles to the surface at the local nanoscale proximity. This implies that spectral properties of sample are mapped with different orientations of the driving dipole (Fig.~\ref{f-toy}(b)). The primary effect in scattering/reflection is smearing spectral features observed on flat samples as experimentally observed on Au-HEA sample sputtered over the black Kapton with high local roughness $\sim 160$~nm (Fig.~\ref{f-blac}). Light scattered from the neighboring nano-hills/slopes/crevices all contribute to the experimentally recorded signal and spectral features characteristic to $n$ and/or $\kappa$ are scrambled. Reflectivity of Au-HEA on black-Kapton was lower by 20-30\% as compared with Si (would be expected $\sim$2 times higher for a metal mirror as in Fig.~\ref{f-mirr}).  

\subsection{Polarisation analysis of tip-scattered/reflected light}

\begin{figure*}[t!]
\centering\includegraphics[width=.95\textwidth]{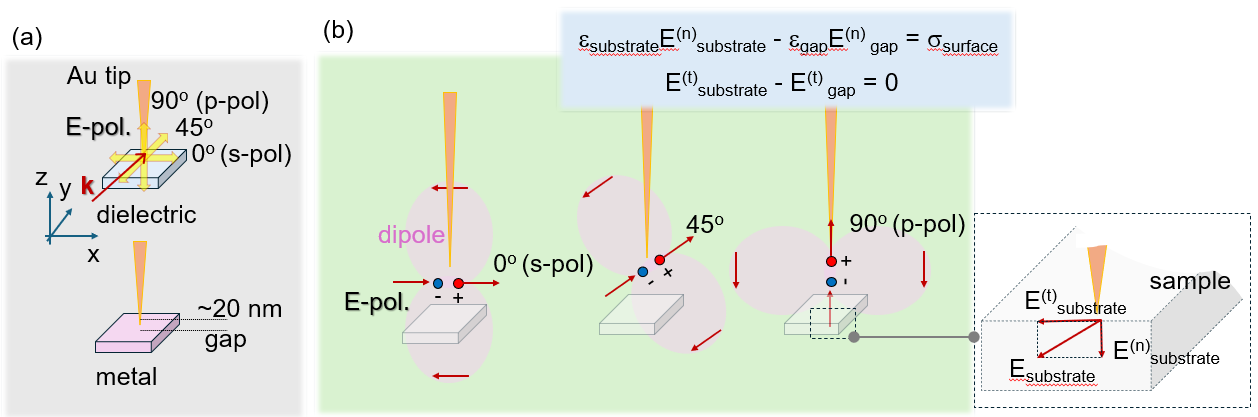}
\caption{\label{f-toy} 
Toy model for FDTD. (a) Nano-needle antenna illuminated at different polarisations angles $\theta = 0^\circ, 45^\circ, 90^\circ$ in the (s,p)-plane of polarisation along the wavevectot $\mathbf{k}$ at slanted angle of incidence; p-pol. corresponds to $90^\circ$ and s-pol. is $0^\circ$. (b) Schematics of the dipole field at different orientations in the nano-gap. The boundary conditions for the normal $(n)$ and tangential $(t)$ components of $E$-field across the substrate-gap interface shown in the inset; $\epsilon\equiv\tilde{n}^2$ is the permittivity and $\sigma_{surface}$ is surface charge density (free and polarisation induced). Geometry: Au nano-cylinder of 20-nm-diameter (length $\sim 200$~nm, nano-gap above sample 20~nm; light is focused with $NA = 0.95$ conditions on the Au-tip. Light is incident on Au-tip along y-direction at an angle of few degrees. Dielectric is \ce{Al2O3}, semiconductor Si, metal Au, wavelength $\sim 1~\mu$m (see discussion of Fig.~\ref{f-ey}). 
   }
\end{figure*}

The $E_z$ normal to surface field is corresponding to the p-pol. E-field in attenuated total reflection (ATR) spectroscopy widely used at long-IR and THz wavelength where absorption in air is considerable. For the ATR mode of IR-micro-spectroscopy, it was demonstrated that anisotropy of absorbance in the sample can be measured using incident polarisation with inclined linear polarisation in the plane of (s,p)-polarisations. The 4-pol. method was demonstrated for non-propagating near-field using crystallised polymer samples~\cite{22nz1047}. This polarisation tomography principle by non-propagating near-field probing absorbance and retardance inside sub-surface of the sample is based on the very same nature of dipole interaction as in the nano-IR. Arguably, the use of polarisation of an azimuthally tilted incident light in (s,p)-plane of light incident onto nano-probe, (sub-)surface anisotropy could be tested (see inset in Fig.~\ref{f-toy}(b)). This conjecture is tested numerically by FDTD calculations in this section. It is expected to follow similar component contributions as for ATR~\cite{21as7632}.

The general expression of the power measured (from the interaction volume in ATR) is the addition of two perpendicular s- and p-polarisations and follows energy conservation~\cite{21as7632}:
\begin{equation}\label{E1}
 P^{(OUT)}(\varphi) = A_s\times E_s^2\cos^2(\varphi+\Delta\vartheta_s) + A_p\times E_p^2\sin^2(\varphi+\Delta\vartheta_p) , 
\end{equation}
\noindent where $A_{s,p}$ are the arbitrary amplitudes accounting for both absorption losses and changes in amplitude due to reflection (from sample), $\Delta\vartheta_{s,p}$ are the phases for s- and p-polarisations, respectively, while the angle $\varphi$ defines the orientation of the linear polarisation in the incident beam: for $\varphi=0^\circ$ only $E_s$ field is present and $\varphi=90^\circ$ corresponds to pure $E_p$ (depolarisation is expected from interaction nano-volume and even for pure s-/p-pol incidence, there could be both contributions in output scattered power in nano-IR). 

In nano-IR, Eqn.~\ref{E1} would describe a change of detected s-/p-pol. components according to the coefficients $A_{s,p}$ accounting for absorption induced changes (dichroism) at incident intensities $|E_{s,p}|^2$ as well as due to phase changes between polarisations $\Delta\vartheta_{s,p}$ (birefringence). Visualisation of Eqn.~\ref{E1} is shown in Fig.~\ref{f-4pol} revealing intricate amplitude changes of detected signal on the affected phase. With a detector with four grid polarisers, the fit to the angular dependence can be extracted as shown in Fig.~\ref{f-4pol}(a). 

Feasibility of orientation dependent intensity control using antennas is demonstrated in nano-polymerisation where nano-gaps and ends of antennas provides highly localised light enhancement, hence, triggers polymerisation sites~\cite{09jpcc1147,09jphc11720}. It was shown in the nanolithography application by direct laser writing that orientation of incident polarisation onto a nano-groove of 10-20~nm in width in a dielectric nano-film promotes nanoscale ablation via enhancement inside the nanogap (nano-groove)~\cite{20lsa41}. This is based on boundary conditions for the normal E-field component to the interface gap-dielectric. The very same dipole light localisation and energy deposition was also used in 3D modification inside dielectrics and semiconductors~\cite{24np799}. All these examples show orientation/direction resolved action of non-propagating near-fields. 

\subsection{Numerical modeling of the polarisation effect}

The fundamentals for nano-IR spectroscopy and mapping established more than 20 years ago and widely used now are based on the analytical dipole based antenna interaction with surfaces. Once permittivity $\epsilon = \epsilon' + i\epsilon" \equiv (n + i\kappa)^2$ is established at the required wavelength and nanoscale position~\cite{Govyad}, the light-matter interaction can be quantitatively described for the incident light field.   

Without analytical rigor to model the actual experimental setting, a numerical model is tested here for Au-needle illuminated at few polarisation angles with dielectric and metallic samples (Fig.~\ref{f-toy}). The FDTD simulations provide precise description of nanoscale light intensity distributions, which can visualise orientations of light matter interactions according to the intensity distributions. This opens a possibility to image anisotropies in absorbance (dichroism) and refractive index (birefringence) as tested in the ATR mode~\cite{22nz1047}. This toy-model approach could provide justification of analytical approach to probe different polarisations of the scattered field, which are related to the local $(n, \kappa)$ or $(\epsilon',\epsilon")$ dependent on the orientation of the intensity in sub-surface of the sample as it was demonstrated in ATR and nano-ablation~\cite{22nz1047,20lsa41}.    

The physics of near-field localisation and polarisation-dependent field distribution is scale invariant in this regime, and therefore independent of the exact wavelength choice. The model was set for 1~$\mu$m wavelength, which is much smaller than the mid-IR wavelengths used in experiments but allowed the use smaller 3D calculation volume. The 20-nm-gap between 20-nm-diameter Au cyllinder and sample is illuminated with Gaussian beam using a $NA = 0.95$ objective lens when sample is of typical materials: Au, Si, \ce{Al2O3} representing metal, semiconductor and dielectric, respectively. The normal (n) and tangential (t) E-field components in sub-surface region of the sample are of interest (Fig.~\ref{f-toy}(b)). They are non-propagating optical near-fields, however, they can be used for measurement of polarisation/orientation dependent anisotropies in absorption~\cite{22nz1047,20lsa41} as discussed above. 

\begin{figure*}[tb]
\centering\includegraphics[angle=0,width=1\textwidth]{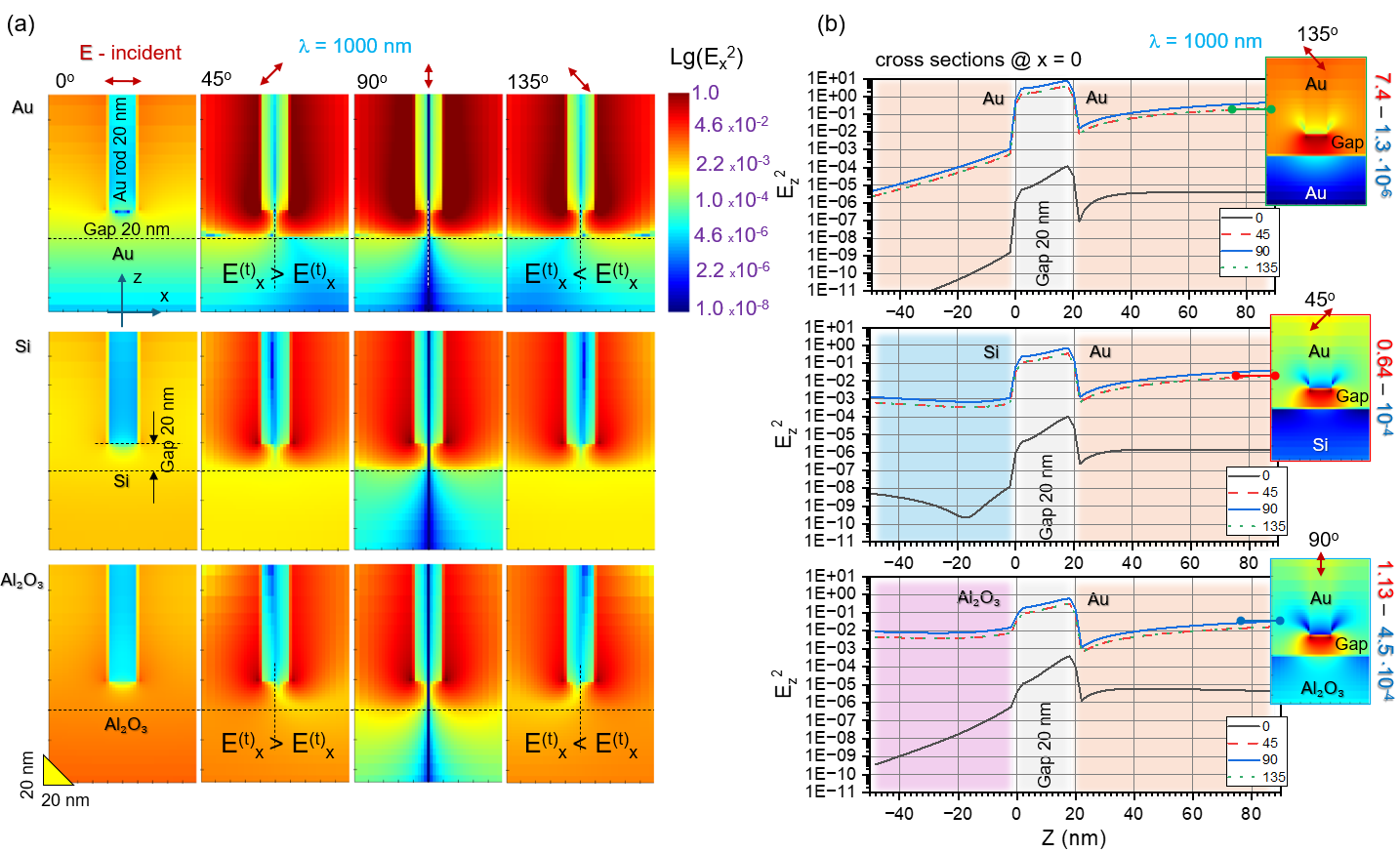}
\caption{\label{f-ey} (a) FDTD calculated intensity in the xz-plane (perpendicular to k-vector), see the model in Fig.~\ref{f-toy}. The intensity component $|E_x|^2$ is plotted on a log-scale. Polarisation of the incident light is rotated by $45^\circ$ from $0^\circ$ to 135$^\circ$ to change ratio and orientation of normal and transverse E-field components $(E_z,E_x)$, respectively. All intensity colormaps for Au, Si, \ce{Al2O3} are normalised to the same range. The transverse E-field components $E^{(t)}_x$ are different on the positive $x>0$ and negative $x<0$ sides for the incident polarisations $45^\circ$ and $135^\circ$ (mostly pronounced in Au). (b) The $E_z^2$ at $x=0$ (center cross section along the beam propagation). This presentation is the most informative since this normal field is symmetric on both sides of z-axis for all incident polarisations. The xz-plane intensity $E_z^2$ maps at selected orientations are shown in insets; intensity is on the linear scale. 
 }.
\end{figure*}

Figure~\ref{f-ey} shows summary of $E^{(t)}\sim E_{x}$ in (a) and $E^{(n)}\sim E_z$ in (b). The (xz)-plane intensity cross sections of $E_x$ component are plotted on the same scale for three used sample materials Au, Si, and \ce{Al2O3}; the scale is logarithmic. A distinguishable difference of $E_y$ components at $x>0$ and $x<0$ sides around the center cross section for the $45^\circ$ vs $135^\circ$ incident polarisations was evident in the case of Au and less pronounced differences were for Si and \ce{Al2O3} (see $E_x$ fields at 632~nm in Fig.~\ref{f-phase-wrapping}). The intensity $E_z^2$ - the normal $(n)$ component - on-axis (z-axis) through the Au-rod is shown in (b) for three samples. There were no asymmetry of $E_y^2$ on two sides of z-axis (this is why the (xz)-plane views are omitted). Selected (xz)-plane cross sections are shown in (b) for the one of the orientations of incident polarisation for illustration. The $E^{(n)}$ is localised inside the air gap as expected from the boundary conditions (see Fig.~\ref{f-toy}(b)). 

Figure~\ref{f-ey} shows that the intensity of the near-field $(E_x + E_z)^2 \equiv (E^{(t)} + E^{(n)})^2$ (in-plane intensity where $(n)$ is p-pol. and $(t)$ is s-pol.) has an azimuthal orientation defined by the orientation of the incident polarisation. A tilted nano-probe geometry was also simulated for both tangential $(t)$ and normal $(n)$ modes (Fig.~\ref{f-angle}) and also confirms possibility to form sub-surface E-field distribution, which is non-symmetric in respect to the center point (tip-to-sample's surface). Tilted tips have already been demonstrated in 3D-AFM~\cite{Zhang19} and could therefore be employed for optical probing.  
One can expect that absorption anisotropy can be probed/measured using tip-based near-field spectroscopy. Such principle was demonstrated in the case on non-propagating ATR. By adding rotation (sample or illumination of nano-tip) mapping of sub-surface anisotropies can be realised similarly to the computer tomography.  

\section{Conclusions and outlook}

Nea-SCOPE image of metallic HEA was carried on a top-surface of sputtered Au-HEA film and its cross-sectional slice throughout the $\sim 1~\mu$m thickness. 
The recorded nano-FTIR spectra of the metallic alloy in the cross-section sample show strong phonon absorption, characteristic of metal oxides. Reflectivity is lower than that of Si and is not very uniform across the examined area of few micrometers in cross section.
When the Au-HEA is deposited on a cover glass substrate, the nano-FTIR reflectivity is high and typical of a metallic sample. The contribution from the metal oxide is reduced. When Au-HEA was deposited on a black Kapton substrate, low reflectivity was observed on high roughness $\sim 150$~nm (min-max) coating of $\sim 1~\mu$m thickness.

One could expect coupling of synchrotron-IR radiation, used for FTIR spectroscopy and mapping, to AFM nano-needle antenna 
will be beneficial to nanoscale IR characterisation due to added modality of polarisation probing: perpendicular to the surface rather parallel as in the standard FTIR microscopy. Development of nano-IR for anisotropy probing at nanoscale via 4-pol. method is the next formidable challenge with analysis tools capable of linking spectroscopy to molecular orientation~\cite{25cbm110573}. The FDTD modeling showed that sub-surface anisotropy in absorption could be mapped with orientation-dependent near-fields from the tip-antenna. 

Augmentation of nano-IR with polarisation sensitive analysis could bring new capabilities for nanoscale charaterisation of quantum emitters and nano-devices. It can be combined with complimentary non-destructive SEM characterisation when metal coatings are not required and UV-C co-illumination is used for generation of secondary electrons and reduce surface charging~\cite{13lpr1049,16aplp021301}. Nanoscale optical and material characterisation can be obtained without destruction of functional nano-devices.

\small\begin{acknowledgments}
S.J. acknowledges support via ARC DP240103231 grant. 
A.S.M.A acknowledges support via ARC DP210103318 grant. It was also supported by the Industrial Transformation Training Centre project IC180100005, 
- ``Surface Engineering for Advanced Materials'' (SEAM). J.M. acknowledges support via JST CREST (Grant No. JPMJCR19I3) and KAKENHI (No.22H02137). M.R. was supported via KAKENHI (Grant No. 22K14200) and SAKIGAKE (Grant No. JPMJPR250E). We are grateful for the vital support from our collaborators in industry, academia, and other organizations, whose contributions were essential to the development and maintenance of SEAM. We are grateful to Jimmy Wales for continuous support. S.J. is grateful to the Nanotechnology Ambassador program at the MCN-ANFF node in 2024-25. Supplementary investigations were performed using synchrotron-based FTIR technique on the Infrared Microspectroscopy (IRM) beamline at the Australian Synchrotron, part of ANSTO, through merit-based beamtime proposals (ID. 20039, 20635, and 22505).  
\end{acknowledgments}

\bibliography{aipsamp}


\appendix
\setcounter{figure}{0}\setcounter{equation}{0}
\setcounter{section}{0}\setcounter{equation}{0}
\makeatletter 
\renewcommand{\thefigure}{A\arabic{figure}}
\renewcommand{\theequation}{A\arabic{equation}}
\renewcommand{\thesection}{A\arabic{section}}

\begin{figure*}[h!]
\centering\includegraphics[width=1\textwidth]{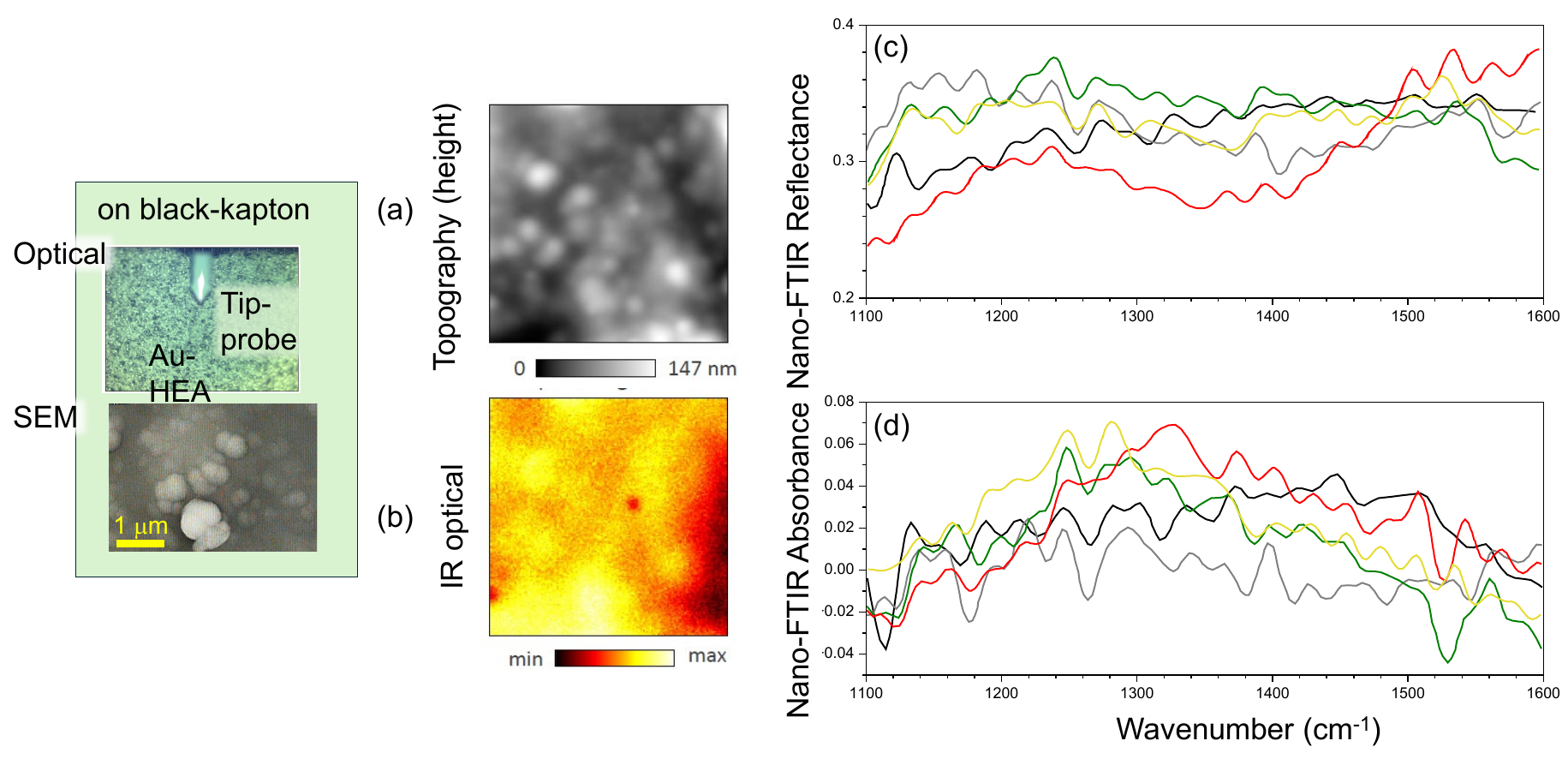}
\caption{\label{f-blac} 
Topography (height) and IR optical maps: $5\times 5$~mm$^2$, $150\times 150$~pixels, 9~ms/pixel. Nano-FTIR reflectance and absorbance at six random locations (normalised to Si). Sample: cover glass coated with Au-HEA. Experiment: 5 averaged interferograms per measurement point; 1024 pixels per interferogram, 10 ms per pixel, spectral resolution 12.5 cm$^{-1}$. Negative $A$ values are the normalisation of Au-HEA to Si result and occurs at the wavelengths where Au-HEA is less lossy as compared to Si}.
\end{figure*}

\section{Nano-IR from nano-rough surface }\label{Sup}

Figure~\ref{f-blac} shows $R$ and $A$ point spectra and six locations on nano-rough ($\sim 150$~nm) surface of $\sim 1~\mu$m-thick Au-HEA coating (see SEM inset and topography map in (a)). Spectra do not represent the Au-HEA, which was clearly mapped when on mirror (4~nm roughness) surface. Most probable reason of spectral changes is contribution from the closest surface features, which become excited with the dipole field (with specific polarisation orientation and amplitude) and their scattered/reflected components. Arguable, polarisation discrimination of the detection could improve spectral finger printing of particular material, e.g., Au-HEA, even on a nano-rough surface.  

\section{Towards polarisation resolved nano-IR}

\begin{figure*}[tb]
\centering\includegraphics[width=1\textwidth]{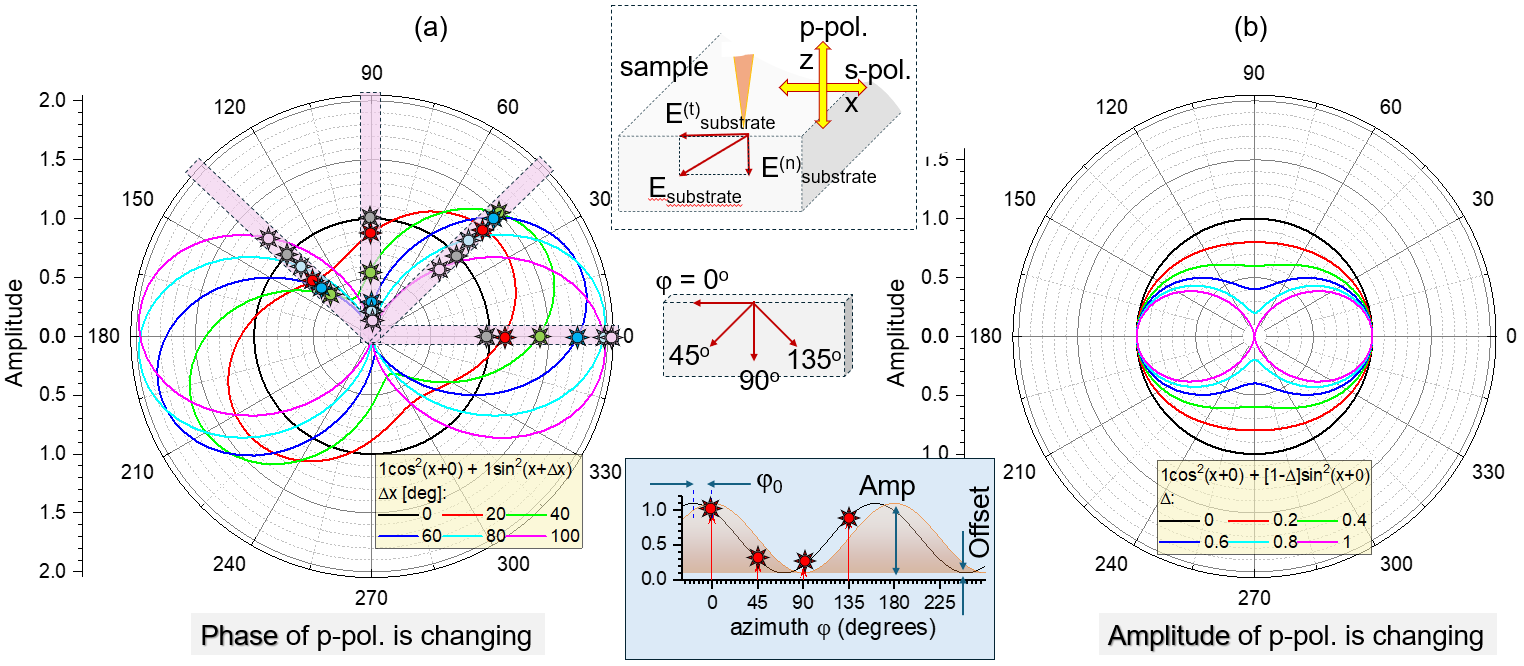}
\caption{\label{f-4pol} 
Visualisation of Eqn.~\ref{E1}: for the incident $E_s = E_p = 1$ and scattered/reflected (detected) light at changed amplitude or phase of the one $E_p$ (y-direction). (a) Only the phase of p-pol. $\Delta\vartheta_p$ is changing in back-scattered intensity/power. (b) Only the amplitude of the p-pol. $A_p$ is changing. Inset shows the sub-surface of sample with normal and tangential (p- and s-pol.) contributions to the scattered/detected field. Rectangular markers in (a) show the intensity/power readout of scattered light at four polarisations of a detector with wire-grid polarisers rotated to each other by 45$^\circ$. The colored dot-stars show expected readout of detector at corresponding $0^\circ, 45^\circ, 90^\circ, 135^\circ$ wire grid polariser orientations. The 4-pol. fit in the bottom inset is by intensity $I = Amp\times\cos^2(\varphi+\varphi_0)+\mathrm{Offset}$, with amplitude $Amp = 1$, offset $\mathrm{Offset} = 0.1$ and the angle of maximum absorbance when $I$ is maximum $\varphi_0 = -19^\circ$. 
   }
\end{figure*}

Complexity of back-scattered/reflected light from the tip of nano-probe is illustrated in Fig.~\ref{f-4pol}, which depicts Eqn.~\ref{E1}. The coefficients $A_s$ and $A_p$ defines how much the amplitude of the incident intensities $|E_s|^2$ and $|E_p|^2$, correspondingly, have changed due to absorption losses inside the sample and are sensitive to the dichroism. The change of the phases between two perpendicular s- and p-pol. due to optical retardance defined by the birefringence is reflected in changes of the phases $\Delta\vartheta_s$ and $\Delta\vartheta_p$. Fig.~\ref{f-4pol}(a) shows that if relative phase between s- and p-pol. is changed, the detected scattered/reflected power/intensity is changing azimuthal orientation and amplitude in the (s,p)-plane. A simpler changes are expected for one amplitude change shown in Fig.~\ref{f-4pol}(b).

The principle of 4-pol. analysis carried out with non-propagating near-field light in ATR geometry was already demonstrated~\cite{22nz1047} experimentally. By changing azimuth of incident light onto the sample/prism in (s,p)-plane, the near-field intensities are probing the sub-surface of the sample at tilted orientations. The back-scattered field into the propagating mode is detected and polarisation analysed. The very same situation is expected for the sample excitation by nano-tip with two intensities normal and transversal to the sample $|E^{(n)}|^2$ (p-pol. in back-scattered field) and $|E^{(t)}|^2$ (s-pol. in back-scattered field) as shown in Fig.~\ref{f-toy}(b) and inset of Fig.~\ref{f-4pol}. These near-fields probe sample will be affected by local absorption and refractive index. The difference of those properties in the perpendicular directions are related to the dichroism and birefringence. The back-scattered light from the nano-probe will deliver this information back into propagating field. This was directly measured in ATR THz experiments~\cite{21as7632}. 

Feasibility to use similar polarisation analysis for nano-probe (as was used for ATR) is based on the same principle of near-field probing the sub-surface. The re-scattered near-field into far-field is then detected. Sensitivity of detected light to absorption and refraction (as well as to their anisotropy) is illustrated in Fig.~\ref{f-4pol}. Experimental realisation would have linearly polarised light incident on the tip at different azimuth $\varphi$ and detected light polarisation analysed with the fit using Eqn.~\ref{E1}. Analysis at the detector can be made by a standard four orientation wire grid polarisers (marker points in Fig.~\ref{f-4pol}(a)). One expected simplification of 4-pol. analysis in the nano-probe case (as compared with ATR) is the use of reference. In nano-IR, it is mirror of Au or Si, while in ATR measurement with and without sample, changes of phases and amplitudes of the back-reflected light occur with complex dependencies including $\pi/2$ phase jumps~\cite{21as7632}. \\

\subsection{Tentative protocol for 4-pol. method in (s,p)-plane}

First, for reference (flat Au, Si), one angle $\varphi_{or} = 0^\circ$ (only s-pol.) incidence is illuminating the nano-probe ($E_s = 1$, $E_p = 0$) and analysis of the back-scattered light is carried out at 4 angles $\varphi_{or} = 0^\circ, 45^\circ, 90^\circ, 135^\circ$. This will retrieve $A_s^{(ref)}(0^\circ), A_p^{(ref)}(0^\circ), \vartheta_s^{(ref)}(0^\circ), \vartheta_p^{(ref)}(0^\circ)$; p-pol. parameters also should be measured due to inherent depolarisation of back-scattered light from nano-volume at the antenna tip. Then, sample is illuminated at the same conditions of $\varphi=0^\circ$ with corresponding set of $E_{s,p}$ and the very same parameters are extracted from the fit by Eqn.~\ref{E1} to the detected back-scattered light: $A_s^{(s)}(0^\circ), A_p^{(s)}(0^\circ), \Delta\vartheta_s^{(s)}(0^\circ), \Delta\vartheta_p^{(s)}(0^\circ)$. This is repeated for $\varphi_{or}=45^\circ, 90^\circ, 135^\circ$ with corresponding set of $E_{s,p}$ for fit by Eqn.~\ref{E1} at detector. The difference between parameters $P\equiv(A_{s,p},\Delta\vartheta_{s,p})$: the change from reference $\Delta P(\varphi_{or}) = P^{s}(\varphi_{or}) - P^{ref}(\varphi_{or})$ for all four parameters at four specific orientation angles $\varphi_{or} = 0^\circ, 45^\circ, 90^\circ, 135^\circ$ (Eqn.~\ref{E1}) is calculated and related to the local absorbance for $A_{s,p}$ and refractive index (real part) for $\vartheta_{s,p}$. Once, the set of parameters $\Delta P(\varphi_{or})$ is measured, it can be fitted with $Amp\times\cos^2(\varphi + \varphi_{0})+\mathrm{Offset}$, where $Amp, \varphi_0, \mathrm{Offset}$ are standard fit parameters for 4-pol. analysis (see bottom inset in Fig.~\ref{f-4pol}); sometimes double angle is used based on trigonometric identity $\cos^2\varphi = \frac{1}{2}(1+\cos2\varphi)$. The $\varphi_0$ shows the local orientation of the specific parameter $\Delta P$ in the (s,p)-plane in sub-surface region xz-plane. With added sample/needle translation along the y-direction, a 3D space mapping for the local absorbance and refractive index can be obtained. The described procedure is multi-step, however, its implementation is straight forward. The most important characteristic for expedient measurements, would be simultaneous acquisition of signal at 4-polarisations, which can be challenging for the FTIR measurements.  

With the retrieved back-scattered (reflected) $A_{s,p}$ and $\Delta\vartheta_{s,p}$, it is possible to apply the description adopted in ellipsometry for the reflectance ratio (for the E-field) as $\rho=\tan\Psi e^{i\Delta}$, where $\tan\Psi = \frac{|r_p|}{|r_s|} $ is the amplitude and $\Delta=\delta_p -\delta_s$ the phase defined by difference of phases for p- and s-pol. For the herein discussed case: $\tan\Psi e^{i\Delta} = \sqrt{\frac{A_p}{A_s}}e^{i(\Delta\vartheta_p - \Delta\vartheta_s )}$. Application important condition for the refractive index determination in ellipsometry takes place at the angle of incidence $\theta_i$ when $\Delta=\pm\pi/2$ is achieved upon reflection from the sample; i.e. the orientation angle of the polarization ellipse is 0 (the minor/major axes of the polarisation ellipse are aligned with s- and p-pol.)~\cite{spie}.  

\subsection{Link to ellipsometry and future prospects of nano-IR}

Considering that 4-pol. method is capable to determine and separate orientational properties of absorption and refractive index, hence anisotropies due to dichroism and birefringence~\cite{19ass127}, this capability could be added to nano-IR probe methods. Another inherent feature of 4-pol. analysis is resolution not limited by the diffraction limit~\cite{19n732,25nse70099}, which can also improve resolution of nano-probe based techniques due to additional information obtained from the orientation fit. The fit by Eqn.~\ref{E1} requires four unknown parameters $A_{s,p}$ and $\Delta\vartheta_{s,p}$, which require at least four independent measurements for non-ambiguous fit. The 4-pol. camera with 4 wire-grids and 4 different orientations provides solution (see 4 color markers defining each particular closed curve of the same color in Fig.~\ref{f-4pol}(a)). Experimental evidence that orientation analysis of subsurface volume is possible by non-propagating near-field in ATR geometry is proven using many-point (orientation angles) analysis of back-reflected light with Eqn.~\ref{E1}~\cite{21as7632}. Sample rotation around the irradiation spot (equivalent to the illumination source rotation) would open orientation (or polarisation) tomography capability currently not existing at the spectral finger printing IR spectral range. 

The generic ellipsometry formula $\rho=\tan\Psi e^{i\Delta}$~\cite{hunt} can be expressed via Euler's equation for the reflection (for field) coefficients: $\frac{r_p}{r_s} = \frac{|r_p|}{|r_s|}\cos\Delta + i\frac{|r_p|}{|r_s|}\sin\Delta$. The magnitude is $\left(\frac{|r_p|}{|r_s|}\right)^2 = \frac{|r_p|}{|r_s|}\cos^2(\delta_p-\delta_s) + \frac{|r_p|}{|r_s|}\sin^2(\delta_p-\delta_s)$ and has an expression resembling Eqn.~\ref{E1}. Further reshaping of equation for the intensity, e.g., p-pol is $R_p\equiv |r_p|^2 = |r_s|\cdot|r_p|\cos^2(\delta_p - \delta_s) + |r_p|\cdot|r_s|\sin^2(\delta_p - \delta_s) \equiv \frac{|r_s|}{|r_p|}\cdot R_p\cos^2(\delta_p - \delta_s) + \frac{|r_p|}{|r_s|}\cdot R_s\sin^2(\delta_p - \delta_s)$. 

\section {Phase wrapping in the calculated phase maps}\label{phase-wrapping}

\begin{figure*}[tb]
\centering\includegraphics[width=1\textwidth]{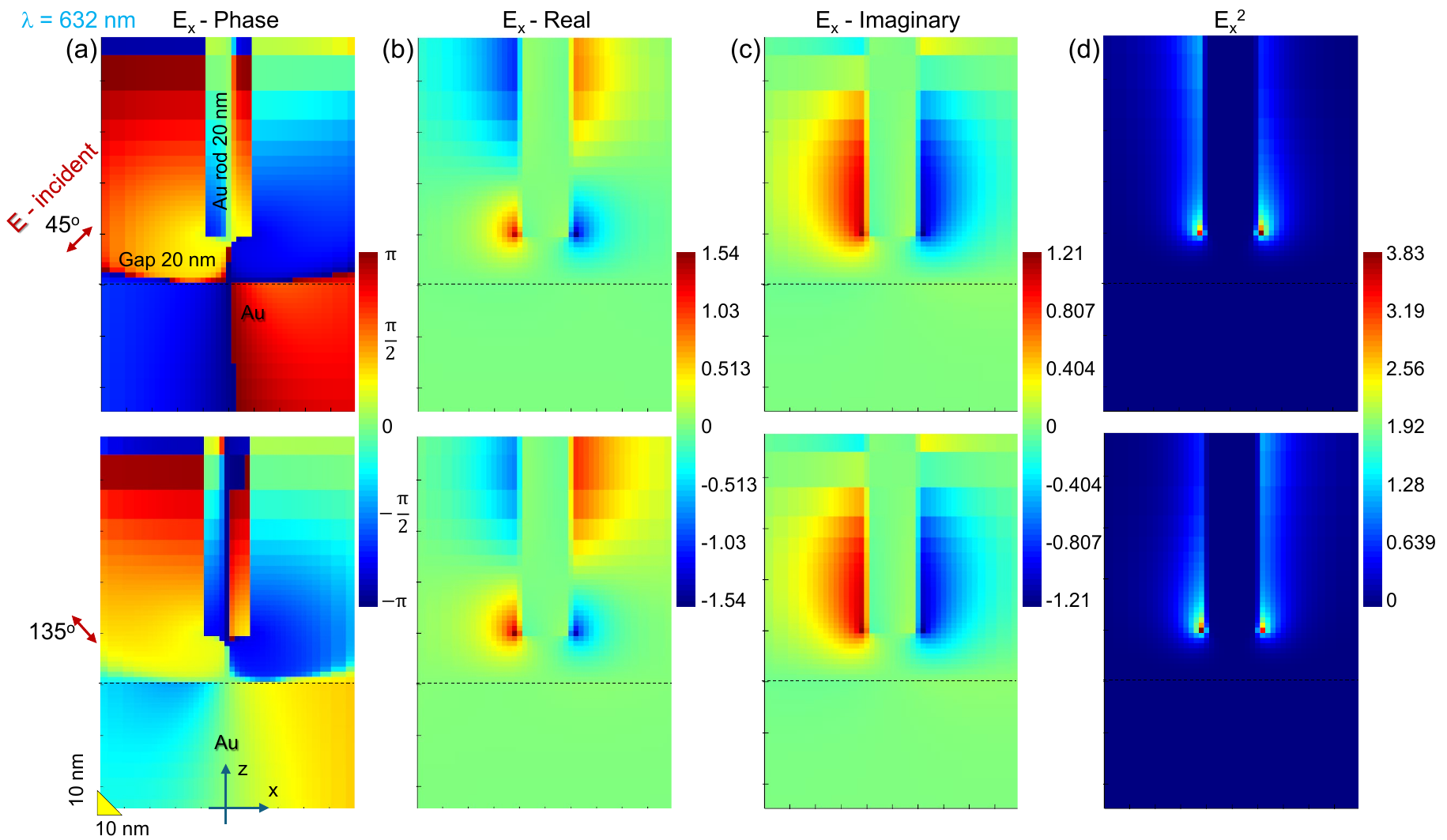}
\caption{\label{f-phase-wrapping} 
Phase (principal branch), real, imaginary and intensity distributions of $E_x$ for $45^o$ (top row) and $135^o$ (bottom row) polarisations at $\lambda$ = 632 nm. The wrapped phase maps are plotted in the principal branch $[-\pi, \pi]$. Although the phase distributions appear visually asymmetric, the corresponding real and imaginary components exhibit clear mirror symmetry with sign inversion. The intensity distributions $|E_x|^2$ remain symmetric between the two cases, confirming that the apparent phase asymmetry originates from $2\pi$ phase wrapping rather than a physical breaking of symmetry. All real and imaginary maps are plotted using identical symmetric color scales; all scales are linear.}
\end{figure*}

The phase of the electric field was calculated as:
\begin{equation}\label{E-A1}
\phi = \mathrm{atan2}\left( \Im\{E\}, \Re\{E\} \right),
\end{equation}
\noindent which is restricted to the principal branch $[-\pi,\pi]$. Figure~\ref{f-phase-wrapping} shows the phase, real, imaginary and intensity distributions of $E_x$ for the $45^\circ$ and $135^\circ$ polarisations (incident onto nano-probe). While the wrapped phase maps appear visually asymmetric between the two cases, the corresponding real and imaginary components exhibit clear mirror symmetry with sign inversion,
$E_{135}(y,z)\approx -E_{45}(-y,z)$, consistent with the expected $\pi$ phase shift between orthogonal polarisations.

The intensity distributions $|E_x|^2$ (tangential (t) component in terms of boundary conditions) remain mirror symmetric for both configurations, demonstrating that the underlying complex field preserves symmetry. The apparent asymmetry observed in the phase maps therefore originates from the principal-branch representation of the phase ($2\pi$ wrapping), rather than from any physical breaking of symmetry. All real and imaginary field maps are plotted (Fig.~\ref{f-phase-wrapping}) using identical symmetric colour scales to ensure a consistent quantitative comparison.

\section {Needle probe at an angle}\label{angle}

\begin{figure*}[tb]
\centering\includegraphics[width=1\textwidth]{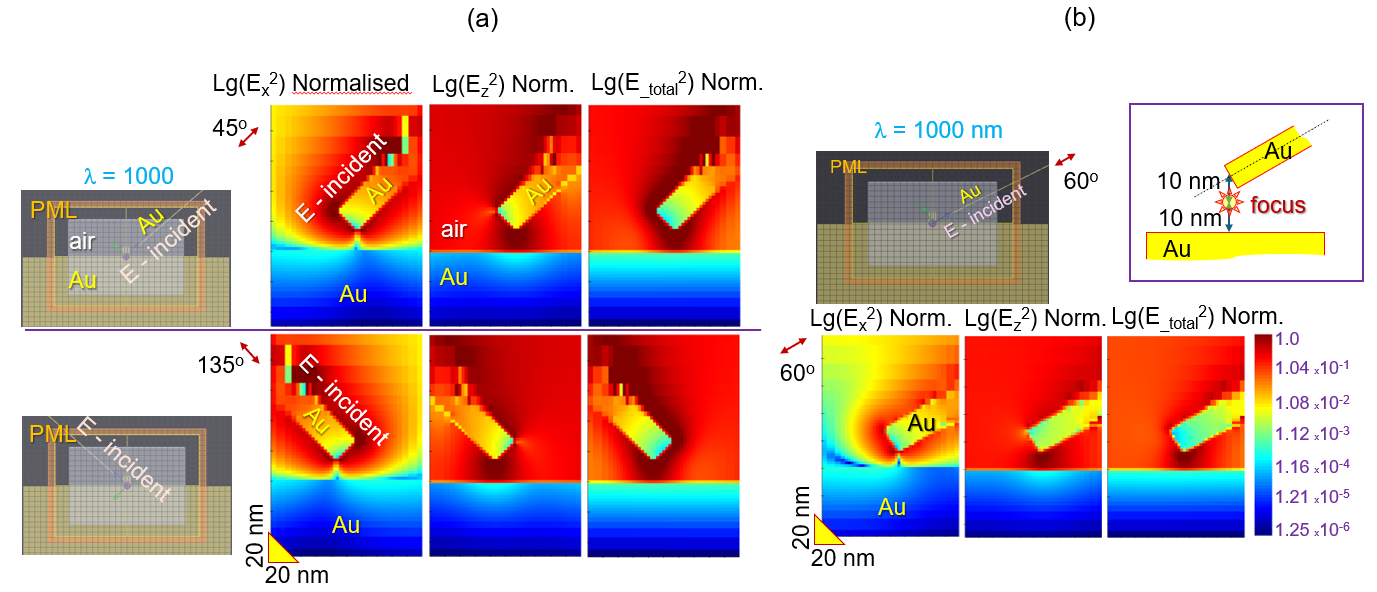}
\caption{\label{f-angle} FDTD calculations for the tangential (t) $E_x$ and normal (n) $E_z$ components for: (a) 45$^\circ$-tilted nano-probe (20-nm-diameter Au rod) and (b) at 45$^\circ$-tilt. Normalised intensity presentation in $\lg$-scale is shown to highlight asymmetry of E-field in subsurface region of the sample (Au). Incident field $E_0 = 1$ is focused onto the gap (like in Fig.~\ref{f-phase-wrapping}). Top-inset in (b) shows focus position in the middle of the gap.
}
\end{figure*}

Figure~\ref{f-angle}(a) shows geometry when the nano-probe is at $\pm 45^\circ$ angle to the normal of the sample. The incident E-field is polarised along the nano-probe. This geometry shows the same tendency of tangential $(t)$ component $E_x$ being asymmetric in respect to the center point on the sample's surface. The normal to the surface E-field component $E_z$ is symmetric with respect to the mid-point on the surface. This geometry shows that the same orientational near-field probing of absorption anisotropies should be possible as with the probe normal to the surface. At 60$^\circ$ angle, qualitatively similar effect for the $(t)$- and $(n)$-components was observed. It is noteworthy that such angle is larger than the magical angle of 54.7$^\circ$ when dichroism is changing from positive to negative in form-birefringent and anisotropic absorbers.  

\end{document}